\documentclass[]{pasj01}
\usepackage{bm}
\begin{document}
\SetRunningHead{3D Structure of Clumpy Outflow}{Kobayashi et al. }

\title{Three-dimensional structure of clumpy outflow from supercritical
accretion flow onto black holes}
\author{%
  Hiroshi \textsc{KOBAYASHI},$^{*}$\altaffilmark{1,2}
  Ken \textsc{OHSUGA},\altaffilmark{2,1}
  Hiroyuki \textsc{R. TAKAHASHI},\altaffilmark{3}
  Tomohisa \textsc{KAWASHIMA},\altaffilmark{2}
  Yuta \textsc{ASAHINA},\altaffilmark{3}
  Shun \textsc{TAKEUCHI},\altaffilmark{4}
  and
  Shin \textsc{MINESHIGE}\altaffilmark{5}}
\altaffiltext{1}{School of Physical Sciences, Graduate University of Advanced Study (SOKENDAI), Shonan Village, Hayama Kanagawa 240-0193}
\altaffiltext{2}{Division of Theoretical Astronomy, National Astronomical Observatory of Japan, 2--21--1 Osawa,Mitaka-shi, Tokyo 181-8588}
\altaffiltext{3}{Center for Computational Astrophysics, National Astronomical Observatory of Japan, 2--21--1 Osawa,Mitaka-shi, Tokyo 181-8588}
\altaffiltext{4}{4-1-10-103, Okusawa, Setagaya, Tokyo 158-0083}
\altaffiltext{5}{Department of Astronomy, Graduate School of Science, Kyoto University, Kitashirakawa,
Oiwake-cho, Sakyo-ku, Kyoto 606-8502}
\email{hiro.kobayashi@nao.ac.jp}
\KeyWords{accretion, accretion disks --- black hole physics --- hydrodynamics --- instabilities --- radiation: dynamics }

\maketitle
\begin{abstract}
We perform global three-dimensional (3D) radiation-hydrodynamic (RHD) simulations of 
outflow from supercritical accretion flow around a $10$ $M_\odot$ black hole. We 
only solve the outflow part, starting
from the axisymmetric 2D simulation data in a nearly steady state 
but with small perturbations in a sinusoidal form being added in the azimuthal direction. 
The mass accretion rate onto the black hole  
is $\sim10^2 L_{\rm E}/c^2$ 
in the underlying 2D simulation data
and the outflow rate is $\sim 10 L_{\rm E}/c^2$
(with $L_{\rm E}$ and $c$ being the Eddington luminosity and speed of light, respectively).
We first confirm the emergence of clumpy outflow,
which was discovered by the 2D RHD simulations, above the photosphere located at 
a few
hundreds of Schwarzschild 
radii 
($r_{\rm S}$) from the central black hole.
As 
prominent 3D features we find that the clumps have 
the 
shape of a torn sheet, 
rather than a cut string,
and that they are rotating around the central black hole with a sub-Keplerian velocity
at a distance of $\sim 10^3$ $r_{\rm S}$ from the center.
The typical clump size is $\sim 30$ $r_{\rm S}$ or less
in the radial direction,
and is more elongated 
in the angular directions, $\sim$ hundreds of $r_{\rm S}$ at most.
The sheet separation ranges from 50 to 150 $r_{\rm S}$.
We expect stochastic time variations when clumps pass 
across the line of the sight of a distant observer.
Variation timescales are estimated to be several seconds for a black hole 
with mass of ten to several tens of $M_\odot$,
in rough agreement with the observations of some 
ultra-luminous X-ray sources.

\end{abstract}
\section{Introduction}
The significant and unique roles of supercritical (or super-Eddington)
accretion flow in astrophysical objects 
have been
recognized quite recently. 
The most remarkable features of 
supercritical accretion flow are intense high-energy radiation 
and massive outflow 
(see 
Chapter
10 of Kato, Fukue, \& Mineshige 2008 and references therein), 
both of which inevitably 
have a great
impact on its 
environment
\citep{2003ApJ...596L..27K, O05, 2011ApJ...736....2O,2014MNRAS.439..503S}.

Supercritical 
accretion is now thought to occur in a variety of
objects. 
Objects
powered by supercritical accretion are sometimes called as 
{\lq\lq}super-Eddington accretors{\rq\rq}.
Good
examples (or candidates) are
ultraluminous
X-ray sources (ULXs; Fabbiano et al. 1989;
Watarai et al. 2001; King et al. 2001; Swartz et al. 2004) and
ultraluminous supersoft
X-ray 
sources
(ULSs; Di Stefano \& Kong 2003; Urquhart \& Soria 2016, Gu et al. 2016; Ogawa et al. 2017).
In addition, some microquasars
(Watarai \& Mineshige 2003;  Done et al. 2007;
Pakull et al. 2010; Vierdayanti et al. 2010)
and 
some of the
narrow-line Seyfert 1 galaxies (e.g., Wang et al. 1999; Mineshige et al. 2000) 
may fall onto this category.
More recently, ULX pulsars 
and
neutron-star systems 
have joined
the group of
super-Eddington accretors
(Bachetti et al. 2014;
\cite{Furst2016};
Kawashima et al. 2016; Israel et al. 2017;
Takahashi \& Ohsuga 2017; Takahashi et al. 2017).
It has also been
suggested that supermassive black holes might have experienced supercritical accretion phase in their formation epoch (see e.g., Inayoshi et al. 2016 and references therein).

Generally speaking, emergence of outflow seems to be ubiquitous
in any accretion 
system.
In particular,
powerful outflows from various types of black hole objects
have now been observationally
established through a number of observations
; see, \citet{1984ApJ...281...90H}, \citet{1997MNRAS.286..513R},
\citet{2000AA...354L..83K}, and \citet{2013MNRAS.430.1102T}
for AGN outflow, and 
\citet{2004AA...418.1061B}, \citet{2005MNRAS.359.1336C}, 
\citet{2009ApJ...695..888U}, and \citet{Miller2015}
for outflow from black hole binaries.
Given this,
it should be interesting to study how they affect their environments. 
Interaction between supermassive black holes and their environments,
so-called AGN feedback, 
is now being studied from various viewpoints 
in relation
to possible co-evolution of supermassive black holes
and their host galaxies (e.g. Silk \& Rees 1998; 
Kormendy \& Ho 2013). 
Since 
the
energy output in either form of radiation or outflow from 
super-Eddington accretors 
is
enormous, 
the feedback effects should be even more important for 
super-Eddington accretors.
In fact, 
ULXs are occasionally accompanied by ULX nebulae,
ionized regions surrounding them, 
although their physical origins are still an open question at the moment
(see, e.g., Feng \& Soria 2011 and references therein).
It might be useful to note here that 
the enormous impacts of
super-Eddington objects, other than black hole objects,
have also been extensively discussed in various astrophysical contexts, 
including luminous blue variables (LBVs), 
Wolf--Rayet 
stars,
classical novae, supernovae and so on
(Davidson \& Humphreys 1997; Nugis \& Lamers 2000;
Shaviv 2000; Smith et al. 2009).

Global multi-dimensional, radiation-hydrodynamics (RHD)
/radiation-magnetohydrodynamics (radiation-MHD) simulations of super-Eddington accretors are 
being carried out rather extensively by many groups 
(e.g., Eggum et al. 1988; Okuda \& Fujita 2000; 
Ohsuga et al. 2005; 2009; S\c{a}dowski et al. 2014; 
McKinney et al. 2014; Fragile et al. 2014; 
Jiang et al. 2014; Hashizume et al. 2015; Takahashi et al. 2016).
Intense radiation from supercritical accretion flows is shown to drive outflow, 
by which significant 
amounts
of mass, momentum, and energy of gas can be blown away 
(Fukue 2004; Takeuchi et al. 2009; Krumholz \& Thompson 2012),
but the nature of the gas outflow is not well understood yet.
By contrast, 
the outgoing radiation part has been rather intensively discussed in
relation to 
observations. 
The most characteristic features 
inherent
to super-Eddington accretors are found in
hard X-ray ranges at around $\sim 10$ keV (e.g., Gladstone et al. 2009).
There is a broad spectral bump, which can be understood in 
terms of Compton up-scattering of soft photons within 
radiation-pressure-driven
outflow
(e.g., Kawashima et al. 2009, 2012;
\cite{2017MNRAS.469.2997N}; Kitaki et al. 2017). 

Here, we address one question: how is matter blown away? 
Part
of the reason for the poor understanding of the outflow resides in 
the need for large-box
simulations with good spatial resolution for clarifying the outflow properties. 
(In the sub-Eddington regime, by contrast, observable intense radiation mainly originates from the 
black hole vicinity so that large simulation boxes are not always
necessary.)
Takeuchi, Ohsuga, and Mineshige
(2013, hereafter T13) were the first to perform 
large-box RHD simulations,
and made a very important discovery in this context:
they found the emergence of clumpy outflow from 
photosphere of 
super-Eddington accretors.
The typical size of the clumps (clouds), 
$\sim 10$ $r_{\rm S}$, corresponds to about one optical depth,
with $r_{\rm S} = 2GM_{\rm BH}/c^{2}$ being the Schwarzschild radius,
where $G$ is the gravitational constant, $M_{\rm BH}$ is the mass of the black hole,
and $c$ the speed of light. 
This fact implies that a sort of radiation hydrodynamic instability is involved with this clump formation,
in addition to the well-known Rayleigh-Taylor instability in the 
radiation-pressure-dominated
atmosphere (see also Takeuchi et al. 2014, hereafter T14).
We wish to stress that the presence of clumpy features was also observationally indicated 
through the significant time variabilities that could be due to porous outflow from luminous objects
(Fabrika 2004; Middleton et al. 2011; Tombesi et al. 2012).
It might be that the BLR clouds 
originate from clumpy outflow
(T13; see similar suggestions by Nicastro 2000; Elitzur \& Ho 2009; Elitzur 2012).

There are a few remarks on the previous simulations (T13).
First, large simulation boxes are essential to find clumpy outflow, since it is expected to appear 
above a photosphere located at 
a few hundreds of $r_{\rm S}$ from a black hole.
Second, magnetic fields do not play a principal role in clump formation, 
since purely RHD simulations can also produce clumpy outflow (T13; T14).
[A photon-bubble instability 
\citep{Arons92,2005ApJ...624..267T}
is not a primary cause of clump formation.]
We thus 
do not
need to incorporate MHD processes.
Finally, the previous 2D study was restricted to the axisymmetric approximation.
Hence, neither 
the three-dimensional (3D)
clump shape nor turbulent motion accompanying non-axisymmetric flow, if
any, could 
be investigated there. 
It is even unclear if clumpy outflow does appear in 3D simulations.
Then, a next question will be: what are the 
3D
shapes of outflow clumps?

In the present study we thus aim 
to clarify
the 3D nature of the clumpy outflow
by performing global, 3D RHD simulations of supercritical outflow.
We will show rather unique 3D features of clumpy outflow that were not anticipated by the 2D simulations. 
The plan of this paper is as follows: In the next section we describe our model, basic equations and assumptions, and numerical procedures.
We then show our results on 3D outflow properties in section 3. 
Comparison with the 2D cases is also examined there.
The final
section is devoted to 
a discussion on the 
observational implications.

\section{Model and calculation methods}
\subsection{Basic equations}  
In the present study, we solve the full set of RHD equations
that take the terms up to the order of $({v}/{c})$:

\begin{equation}
  \frac{\partial \rho}{\partial t} 
  + \bm{\nabla} \cdot (\rho \bm{v})
  = 0,
\label{eq1}
\end{equation}
is the continuity equation,
\begin{equation}
  \frac{\partial}{\partial t} (\rho \bm{v})
  + \bm{\nabla} \cdot (\rho \bm{v} \bm{v} + p_{\rm gas} 
  \bm{I})
  =  - \rho \bm{\nabla} \psi_{\rm PN}
  - \bm{S_1},
  \label{eq2}
\end{equation}
is the momentum equation of gas,
\begin{equation}
  \frac{\partial E_{\rm gas}}{\partial t} 
  + \bm{\nabla} \cdot (E_{\rm gas} + p_{\rm gas}) \bm{v}
  = - \rho \bm{v} \cdot \bm{\nabla} \psi_{\rm PN}
  - c S_0,
  \label{eq3}
\end{equation}
is the energy equation of gas,
\begin{equation}
    \frac{1}{c^{2}}\frac{\partial \bm{F_{\rm rad}}}{\partial t} 
    + \bm{\nabla} \cdot {\bf{P_{\rm rad}}}
    =  \bm{S_1},
\label{eq4}
\end{equation}
is the momentum equation of radiation, and
\begin{equation}
  \frac{\partial E_{\rm rad}}{\partial t} 
  + \bm{\nabla} \cdot \bm F_{\rm rad}
  = c S_0,
  \label{eq5}
\end{equation}
is the energy equation of radiation;
the source terms
in the momentum and energy equations
are explicitly written as
\begin{eqnarray}
  \bm{S}_1 &=& \rho\kappa_{\rm ff}\frac{\bm{v}}{c}
  \left(\frac{4 \pi B}{c}-E_{\rm rad}\right)\nonumber \\
  \; &-& \rho(\kappa_{\rm ff}+\kappa_{\rm es})\frac{1}{c}
     [\bm{F_{\rm rad}} - (\bm{v} E_{\rm rad} + \bm{v} \cdot {\bf{P_{\rm rad}}})],
     \label{eq7}
\end{eqnarray}
and
\begin{eqnarray}
  S_0 &=& \rho\kappa_{\rm ff} 
  \left(\frac{4 \pi B}{c}-E_{\rm rad} \right) \nonumber \\
  \; &+& \rho\left(\kappa_{\rm ff}-\kappa_{\rm es}\right)\frac{\bm{v}}{c^2}\cdot
     [\bm{F_{\rm rad}} - (\bm{v} E_{\rm rad} + \bm{v} \cdot {\bf{P_{\rm rad}}})].
     \label{eq6}
\end{eqnarray}

Here, $\rho$ is the matter density,
$\bm v$ is the flow velocity,
$ E_{\rm gas} \equiv e_{\rm gas} + \rho \bm{v}^2 / 2$ is the total energy of the gas 
(with $e_{\rm gas}$ being the internal energy density of the gas), 
$p_{\rm{\rm gas}}$ is the gas pressure, 
$B = \sigma T_{\rm{\rm gas}}^4 / \pi$ is the blackbody intensity
(with $\sigma$ and $T_{\rm gas}$ being the Stefan-Boltzmann constant and the temperature of the gas, respectively),  
$\bm{I}$ is the unit matrix,
$E_{\rm{rad}}$ is the radiation energy density, 
$\bm{F}_{\rm rad}$ is the radiative flux vector,
and ${\bf P}_{\rm rad}$ is the radiation pressure tensor.
For simplicity, we adopt the gray (frequency-integrated) approximation 
for the radiation terms.
Neither self-gravity nor magnetic fields are taken into account.
We incorporate general relativistic effects by adopting the pseudo-Newtonian potential, 
$\psi_{\rm PN} = -GM_{\rm BH}/(r-r_{\rm S})$, where
$r$ is the distance from the origin
\citep{PW80}.
We consider the electron scattering opacity, $\kappa_{\rm es}$, and the
Rosseland mean free--free absorption opacity, $\kappa_{\rm ff}$,
\begin{equation}
  \kappa_{\rm es} = \sigma_{\rm T}m_{\rm p}^{-1},
\end{equation}
and
\begin{equation}
  \kappa_{\rm ff} =1.7 \times 10^{-25} m_{\rm p}^{-2} \rho T_{\rm gas}^{-7/2}~{\rm cm}^2{\rm g}^{-1},
\end{equation}
where $m_{\rm p}$ is the proton mass and $\sigma_{\rm T}$ is the Thomson scattering cross-section.

The set of equations (\ref{eq1})--(\ref{eq7}) can be closed by using an ideal gas equation of state, 
\begin{equation}
  p_{\rm gas} = (\gamma - 1) e_{\rm gas} = \frac{\rho k_{\rm B} T_{\rm gas}}{\mu m_{\rm p}},
\end{equation} 
and by adopting the M1-closure, which gives the radiation pressure tensor as a function of the radiation energy density as well as the radiative flux 
\citep{1984JQSRT..31..149L}.
Here, $\gamma = 5/3$ is the specific-heat ratio, $k_{\rm B}$ is the Boltzmann constant, and $\mu=0.5$ is the mean molecular weight.
Throughout the present study, we employ $M_{\rm BH}=10$ $M_\odot$.

\subsection{Numerical procedures}
\begin{figure}
  \begin{center}
    \includegraphics[width=\linewidth]{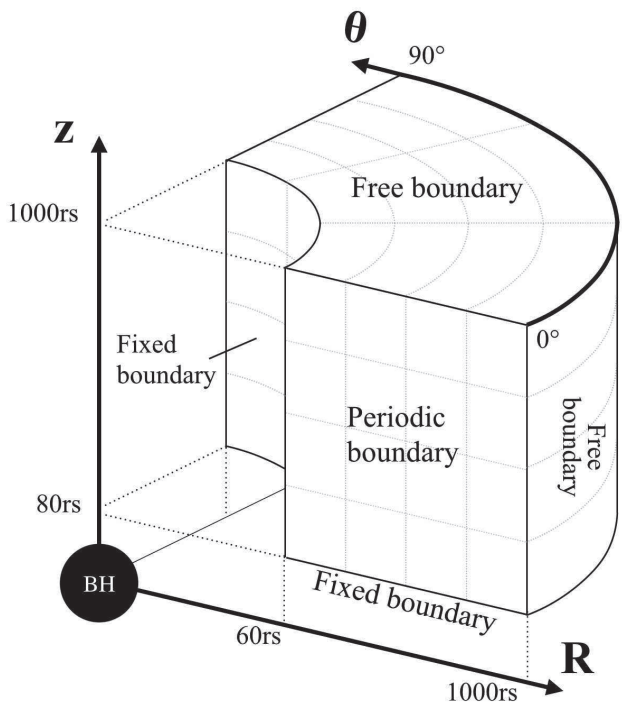}
  \end{center}
  \caption{Density contours used as the initial condition for the 3D simulations (see text).} 
  \label{init}
\end{figure}

In the present study
we use cylindrical coordinates $(R, \theta, z)$, 
where $R$ is the radial distance from the rotation ($z$-) axis,
$\theta$ is the azimuthal angle, 
and $z$ is the vertical coordinate. 
We first performed a 2D simulation (same as T13, except for the size of
the computational box) in order to produce the initial conditions
for the 3D simulation. In the 2D simulation,
computational domain extends from the vicinity of the black hole
($\sim$$2 r_{\rm S}$) to the outflow region of $10^3 r_{\rm S}$. 
A quasi-steady disk, of which the mass accretion rate onto the black
hole ($\dot{M}_{\rm acc}$) 
is $\simeq 10^2 L_{\rm E}/c^2$ and the radiation luminosity
is $\sim L_{\rm E}$, forms in a few sec after the start of the
simulation,
where $L_{\rm E}$ is the Eddington luminosity.
We adopt part of the 2D data at 5 s, but with perturbations being added afterwards
(see below) as the initial condition of the 3D simulation. 
To clarify the initial state, we show the cross-sectional view of the
initial density distribution on the $R$-$z$ plane in Figure \ref{init}.
We have confirmed that the 3D results do not appreciably 
alter, even if we adopt the 2D data at a different time (8 sec, for instance) 
as the initial data for the 3D simulation.

Since the original 2D data obtained above is axisymmetric
($\partial/\partial \theta =0$),
we provide small perturbations as follows:
\begin{equation}
 q (R, \theta, z) = q_0 (R, z) \times [1 + 0.10 \sin(4\theta)],
\label{pert}
\end{equation}
where $q_0$ represents the physical quantities of the original 2D data 
and $q$ is the initial values used in the present 3D simulations. 
We confirm that our results do not appreciably change, even if
we employ random numbers between $-1$ and $1$ instead of
$\sin(4\theta)$ at each grid point. 

\begin{figure}
  \begin{center}
    \includegraphics[width=\linewidth]{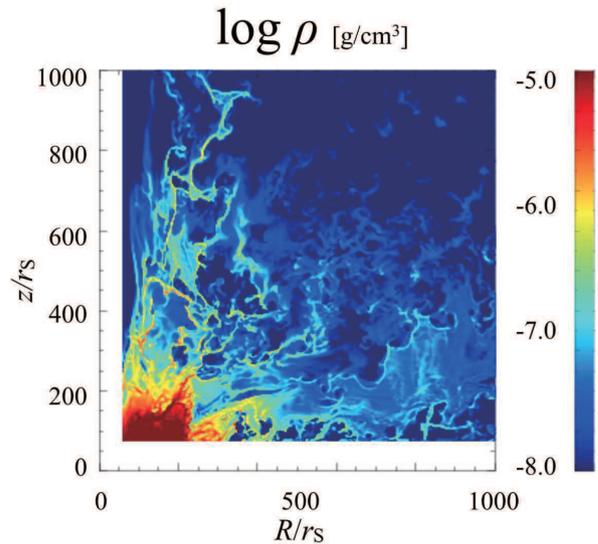}
  \end{center}
  \caption{Geometry of the calculation box and the adopted boundary conditions.} 
\label{sim-box}
\end{figure}

The computational domain of the 3D simulation is 
restricted to
$60 \leq R/r_{\rm S} \leq 10^3$, $0 \leq \theta \leq 90^\circ$,
and $80 \leq z/r_{\rm S} < 10^3$, to reduce the numerical cost. 
At the inner boundary at $R=60 r_{\rm S}$ or at the lower boundary at $z=80 r_{\rm S}$ we keep 
the same values for the physical quantities (such as matter density, temperature, velocity 
vectors, radiation energy density, and radiative flux) at all later times in the 3D simulation.
That is, we employ
fixed boundary conditions at the inner and lower boundaries.
This implies that a steady disk with 
$\dot{M}_{\rm acc} \sim 10^2 L_{\rm E}/c^2$ 
is postulated to exist just inside and below the simulation box
throughout the 3D simulation. 
We adopt free boundary conditions
at the outer and upper boundaries
($R = 10^3$ $r_{\rm S}$ and $z = 10^3$ $r_{\rm S}$).
If the $R$-component ($z$-component) of the velocity is negative, it is
automatically set to be zero at the outer (upper) boundary.
Thus, matter can go out freely but not enter through the boundaries.
The periodic boundary condition is used in the $\theta$-direction so that
the mass, momentum, energy, and radiation across the boundary at $\theta=0^\circ$ 
are the same as those across the boundary at $\theta=90^\circ$.
The simulation box and adopted boundary conditions are summarized in
Figure \ref{sim-box}.
The grid spacing is $\Delta R = \Delta z = 4.0$ $r_{\rm S}$ and
$\Delta\theta = 0.9^\circ$, respectively.
Here we note that the numerical resolution somewhat affects the size of
clumps (discussed later).

Our total simulation time is 12 s, which is much longer than the wind crossing time;
$10^3 r_{\rm S}/v_r \sim$ 1.0 s for the black hole mass of 10 $M_\odot$ 
and the radial wind velocity of $v_r \sim 0.1 c$.
We thus conclude that we finally achieve a quasi-steady structure of the clumpy outflow.

For comparison purposes,
we also perform 2D axisymmetric simulations
of which the computational domain and 
the grid spacing are the same as those of the 3D simulations
($60 \leq R/r_{\rm S} \leq 10^3$, $80 \leq z/r_{\rm S} < 10^3$,
and $\Delta R = \Delta z = 4.0$ $r_{\rm S}$).
The physical quantities of the original 2D data, $q_0$,
are employed as the initial values. As is the case
with the 3D simulations, we use free boundary conditions at 
the outer and upper boundaries, 
and fixed boundary conditions at the inner and lower boundaries.

Our numerical code is an extension of CANS+
\citep{2016arXiv161101775M},
which is a high-resolution MHD simulation code package
developed by Chiba University.
(The magnetic fields are set to be zero in our present simulations).
Hydrodynamic terms are solved by CANS+, and the method for solving
radiation terms is basically same as \citet{2013ApJ...772..127T}.
Here we note that,
while they solve fully special relativistic equations,
we treat the RHD equations to $\cal{O}$$(v/c)$
in the present simulations.

\section{Properties of Clumpy Outflows}\label{sec:figures}
\subsection{Overall structure}
\begin{figure*}
  \begin{center}
    \includegraphics[width=\linewidth]{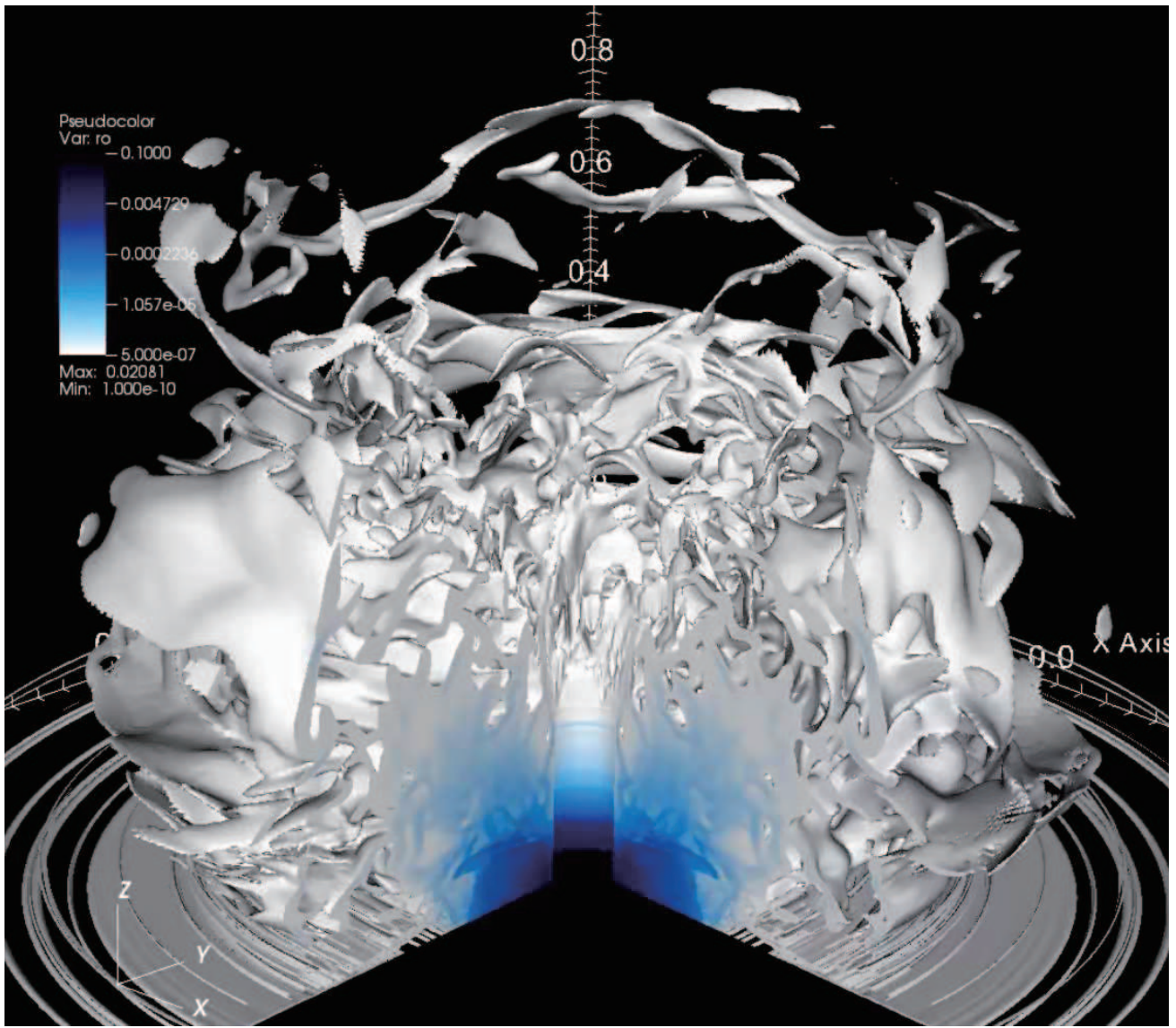}
  \end{center}
  \caption{Bird's eye view of the 3D matter density structure of 
    an outflow 
    from a supercritical accretion flow around a black hole 
    with $M=10$ $M_\odot$ at the elapsed time of $t = 2.5$ s. 
    Here, we only display the regions where matter density is higher than $5\times 10^{-7}$ g cm$^{-3}$.
    The units of each axis are $10^3 r_{\rm S}$ (i.e., the size of the calculation box).
    The color (silver--blue) represents matter density (see the upper-left corner for the color scale); 
    denser regions (indicated by the blue color) are found within the inner inflow region, 
    while less dense regions (indicated by the silver color) are mostly the outflow region.
  } 
  \label{Bird-eye}
\end{figure*}

We first show the overall 3D structure of the outflow gas in Figure \ref{Bird-eye}. This figure displays a bird's-eye view of the matter density distribution for the regions with density being higher than $5 \times 10^{-7}$ g cm$^{-3}$. 
The elapsed time is 2.5 s after the start of our 3D simulation (i.e.,
in a relatively early phase).
We see multiple sheet-like structures in the outflow region (above $z\sim 400$ $r_{\rm S}$). If we have a closer look at each sheet, we notice that its shape is neither an ellipsoid nor a thin string (like spaghetti or linguine), but is more like a flattened string or a torn sheet (like cut fettuccine or lasagna). This feature has been made clear for the first time by the current 3D simulation, and was not anticipated by the 2D simulations.

\begin{figure*}
  \begin{center}
    \includegraphics[width=16cm]{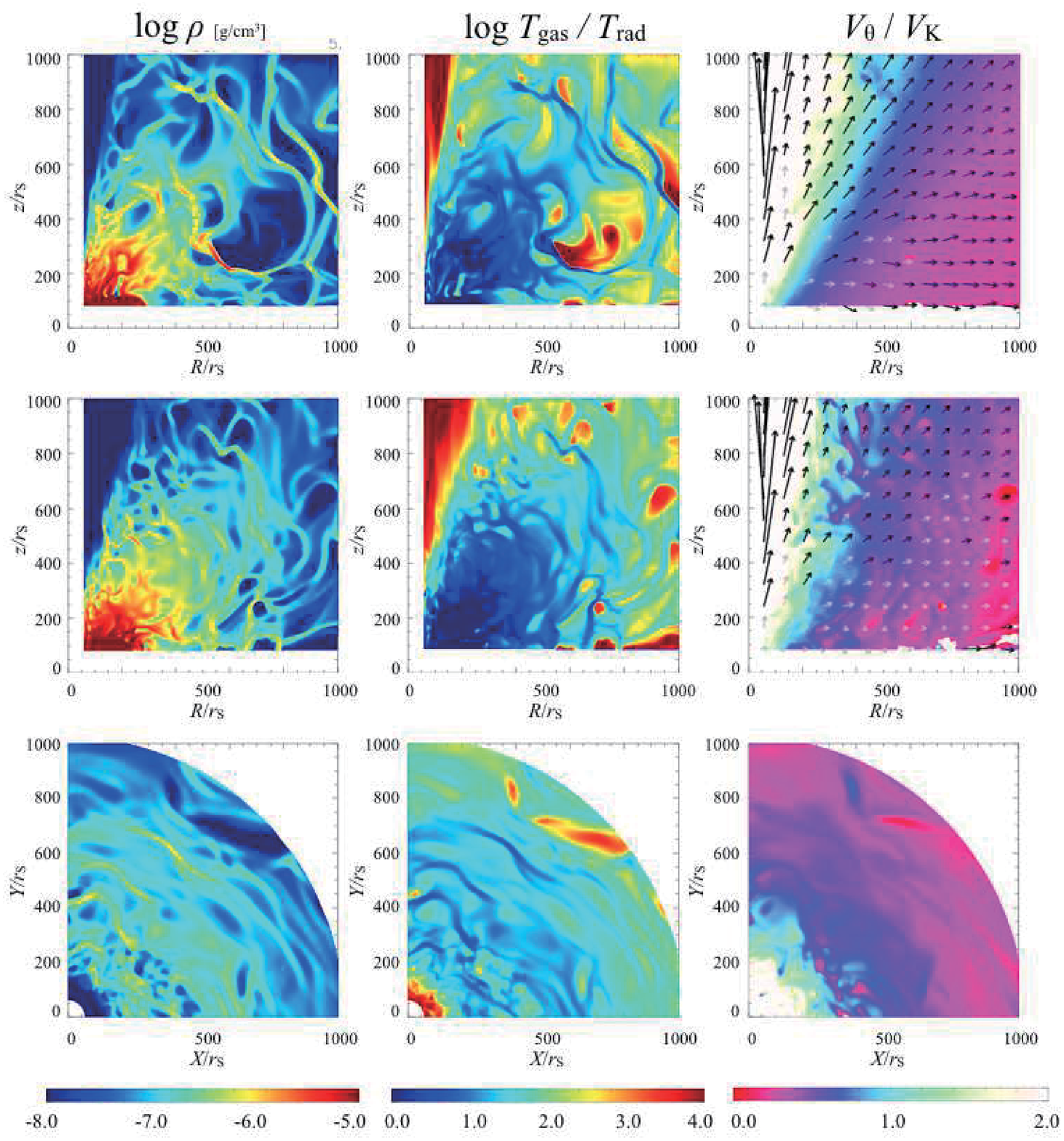}
  \end{center}
  \caption{
Summary of the 3D simulation results but in two-dimensional planes in comparison with the 2D results. 
From the top to the bottom panels, we plot the cross-sectional views of the flow structure in the $R$--$z$ plane calculated by the 2D simulation (upper panels), 
those on the $R$--$z$ plane at $\theta = 45^\circ$ calculated by the 3D simulations (middle panels), and
those on the $X$--$Y$ (or $R\cos\theta$--$R\sin\theta$) plane at $z = 600$ $r_{\rm S}$ by the 3D simulations, respectively.
At each row, we plot the matter density contours (left panels), the contours of the temperature ratio
$T_{\rm gas}/T_{\rm rad}$ (middle panels), and the toroidal velocity fields 
$V_{\theta}$ overlaid with the  poloidal velocity vectors (right panels), respectively. 
The toroidal velocity in the right panels is normalized by the Keplerian orbital velocity, $V_{\rm K}$ 
(see text for the definition). Color scales are indicated below the bottom panels.
The simulation parameters are the same as those in Figure \ref{Bird-eye}
except for the elapsed time, which is 5 s here (i.e., in the quasi-steady outflow state). 
Note the clumpy structures seen in all the matter density contours.
  }
  \label{2Dplots}
  \end{figure*}

Figure \ref{2Dplots} shows a set of 2D, cross-sectional views of the spatial distribution 
of matter density (left panels), temperature ratio (middle), and velocity fields (right) 
of the 2D and 3D simulation results
at the elapsed time of 5.0 s.
From the top to the bottom panels we plot the structure on the $R$-$z$ plane 
calculated by the 2D axisymmetric simulations,
that on the $R$-$z$ plane (or constant $\theta$ plane) by the 3D simulation,
and that on the $X$-$Y$ plane (constant $z$ plane), respectively.
Here, we define the Cartesian coordinates as $(X,Y) \equiv (R\cos\theta, R\sin\theta)$
for a fixed height, $z$.

It is curious to see if there are qualitative differences between the 2D and 3D results. It is obvious that the 2D simulations miss the non-axisymmetric structure, but how about the cases seen in the $R$-$z$ plane?

Let us first focus on the matter density contours (left three panels).
We recognize a clumpy density structure above a certain height ($z >$ several hundreds of $r_{\rm S}$) 
not only in the $R$-$z$ plane but also in the $R$-$\theta$ planes.
It is striking to note that the density contours on the $R$-$z$ planes are amazingly similar among 2D and 3D simulations. This justifies the 2D simulation results.
However, the width of each clump looks somewhat thinner in the 3D calculations. 
This point will be quantified later based on autocorrelation function (ACF) analysis. 
Further, the clumpy structure tends to be weakened in the time-averaged density contours
(not shown, though), meaning that the clumpy structure is not fixed but is fluctuating a lot.
We will show in the next subsection that the clumpy feature is smoothed
by taking the azimuthal average
(again based on the ACF analysis).

It is interesting to examine the correlation of matter-related quantities and radiation-related ones by inspection of the left and middle panels. Apparently they look similar, except for the colors.
This feature was already pointed out by T13, which claimed
anti-correlation between the matter density and the ratio of gas temperature, $T_{\rm gas}$, to the radiation temperature, $T_{\rm rad}$. To be more precise, we notice that the clumpy structure (with higher density seen above $\sim 500$ $r_{\rm S}$) appears in the region where the temperature ratio is smaller;
e.g., the yellow clump (dense region) extending from
$(R, z)=(600$ $r_{\rm S}, 850$ $r_{\rm S})$ to $(750$ $r_{\rm S}, 550$ $r_{\rm S})$ in the middle-left panels 
corresponds to the blue region (where $T_{\rm gas} \sim T_{\rm rad}$) in the central one.
This is consistent with the 2D results reported by T13 (see their Figure 1).
A similar sort of anti-correlation is found in the $R$-$\theta$ plane (see the bottom-left and -middle panels).

Next we compare the velocity fields of the 2D and 3D models. 
We plot in the right panels of Figure \ref{2Dplots}
the azimuthal velocity $(V_\theta)$ normalized by the Keplerian orbital velocity, $V_{\rm K}$, defined as the azimuthal velocity with which matter can rotate on a circular orbit around the $z$-axis at a fixed $z$:
\begin{equation}
    V_{\rm K} \equiv \sqrt{\frac{GM_{\rm BH}R^2}{r^3}},
\label{Kepler}
\end{equation}
with $r \equiv \sqrt{R^2+z^2}$.
The black arrows represent the region where the gas velocity exceeds
the escape velocity,
while the white arrows indicate the region where the velocity is less.

We see a smoother $V_\theta$ distribution in the 2D simulations compared with that in the 3D. 
This is particularly true in the interface between the inflow region (with blue color); 
it is nearly straight in the 2D results (right panel), whereas it is not in the 3D result. 
This is due to significant turbulent motion arising in the interface, producing significant velocity fluctuations. 
In the region around the $z$-axis, a jet with super-Keplerian rotation velocity is launched. That is, 
the jet material tends to go outward, but such an expanding motion is not observed. 
This indicates that the jet material is confined by an external pressure asserted by matter outflow.

It is interesting to note from the bottom-right panel that the toroidal velocity is mostly sub-Keplerian
at large $R (> 400$ $r_{\rm S})$. The rotating gas there is, however, going outward due not to 
the centrifugal force but to the outward radiation-pressure force. 
We understand that material is being blown away, keeping its angular momentum,
since the radiation force in the $\theta$ direction is too small to change the angular momentum.

It is also important to note that significant sub-Keplerian rotation means a slow rotation speed,
which in turn leads to long flux
variation timescales caused by obscuration of central light by moving clumps.
Simple timescale calculations made based on the assumption of Keplerian rotation could be grossly
underestimated. This point will be discussed in the final section.

\subsection{Autocorrelation analysis}

\begin{figure}
  \begin{center}
    \includegraphics[width=\linewidth]{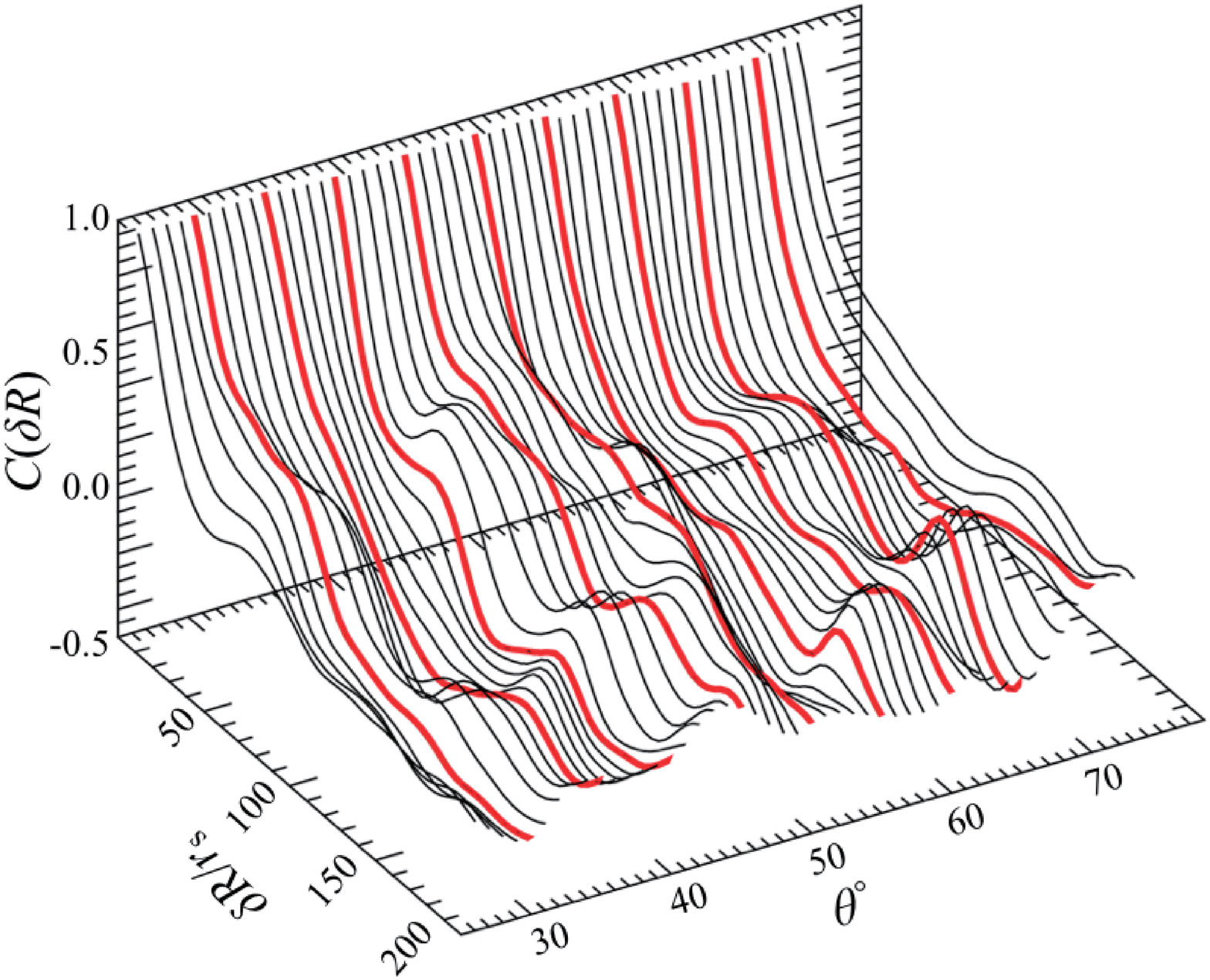}
  \end{center}
 \caption{Autocorrelation functions of the matter density distribution displayed in the 
lower-left 
 panel in figure \ref{2Dplots} as functions of $\delta R$ for a variety of azimuthal angles
 $\theta$.
 We colored some lines red for visual clarity.
  }
  \label{ACF}
\end{figure}

To examine the statistical properties of clumps, it is useful to calculate ACFs (see T13). 
The ACF as a function of the 
radial and azimuthal interval 
($\delta l = \delta R, \delta \theta$)
for example, is calculated by
\begin{equation}
 C(\delta l) \equiv \frac{\displaystyle\sum_{k=0}^{N -L-1}
  (\rho_{k+L}-\bar{\rho})(\rho_{k}-\bar{\rho})}
  {\displaystyle\sum_{k=0}^{N-1}(\rho_k-\bar{\rho})^2},
  \label{ACF-eq}
\end{equation}
where $\bar{\rho}$ is the average density,
\begin{equation}
  \bar{\rho} \equiv \frac{1}{N }\sum_{k=0}^{N-1}\rho_k,
\label{rhobar}
\end{equation}
$N$ is the total number of grid points 
in the radial ($N=235$) and azimuthal ($N=100$) directions,
the subscript $k$ represents the grid number, 
and $L$ is an integer related to 
$\delta l$ as  
$\delta R = l \times \Delta R$
and 
$\delta \theta = l \times \Delta \theta$, 
where we took a constant grid spacing : 
$\Delta R = 4.0r_\mathrm{S}$ and $\Delta \theta = 0.9^\circ$.

\begin{figure}
  \begin{center}
    \includegraphics[width=0.8\linewidth]{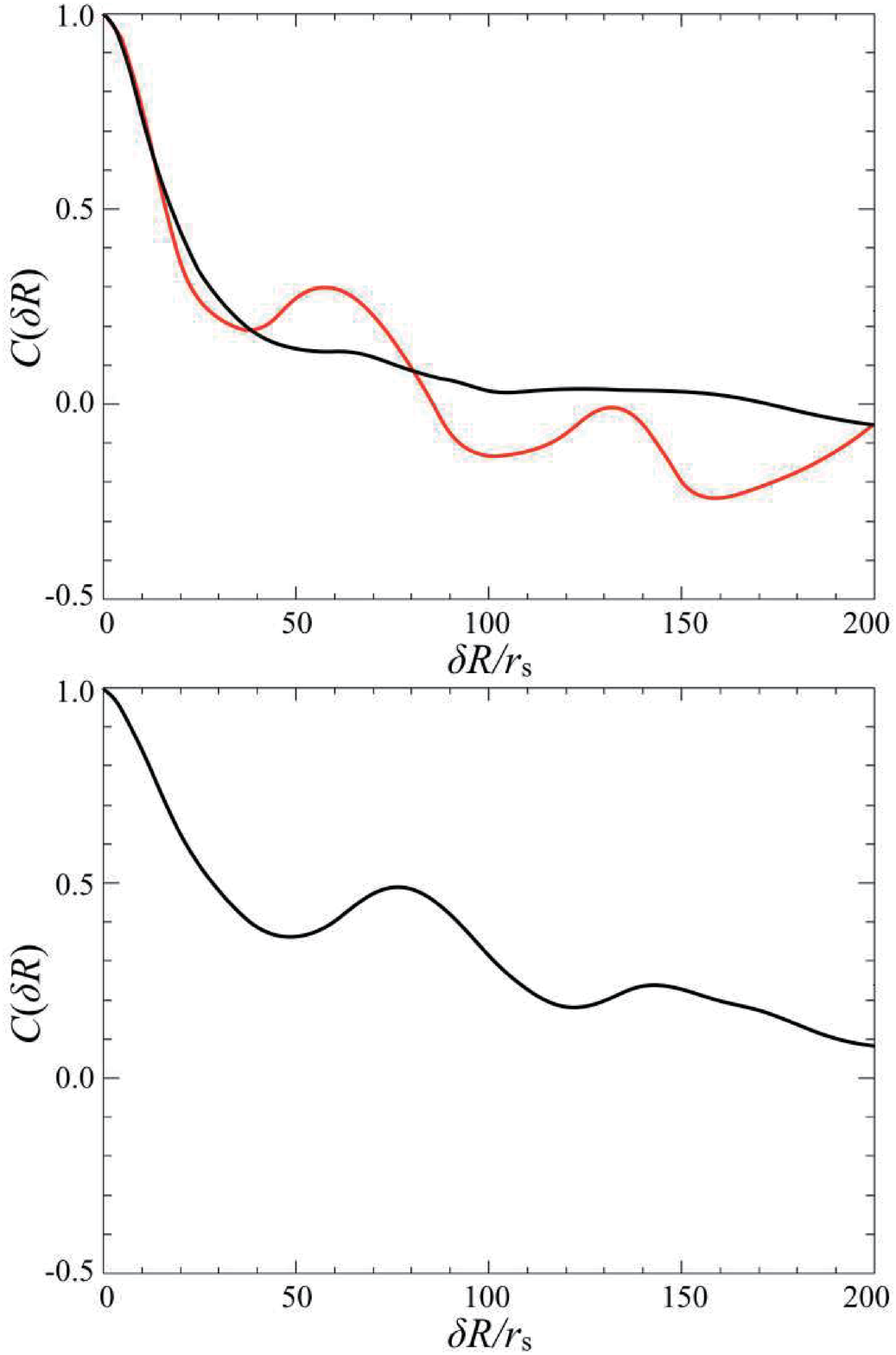}
  \end{center}
  \caption{Comparisons of the ACFs in the 2D and 3D models.
(Upper panel) The ACF as a function of $\delta R$ at fixed $\theta = 45^\circ$ (by the red line)
and the azimuthally averaged ACF (by the black line) obtained by the 3D simulations.
(Lower panel) The same but for the 2D simulations.  
  }
  \label{ACF-R}
\end{figure}

Figure \ref{ACF} illustrates the ACFs calculated based on the matter density data on the $X$-$Y$ plane 
(shown in the middle-left panel of figure \ref{2Dplots})
as functions of $\delta R/r_{\rm S}$
at various angles of $\theta$. 
Here, we fix $z = 600$ $r_{\rm S}$.
See also the red line in the upper panel of Figure \ref{ACF-R}, which is the one-dimensional
plot of $C(\delta R)$ at a fixed $(\theta, z) = (45^\circ, 600$ $r_{\rm S})$.

From these ACF profiles, we can extract the typical size of clumps and typical interval between neighboring clumps in the following way:
First, the width of the primary peak (at around $\delta R=0$)
represents the typical size of each clump.
We find similar slopes of each ACF near $\delta R\sim 0$ in Figure \ref{ACF},
and hence understand that the half-width of the clump 
is $\ell_{\rm cl}^r/2 \sim 15$ $r_{\rm S}$ regardless of the azimuthal angles.
Second, the interval between the primary peak and the secondary one at 
$\delta R = 50 - 150$ $r_{\rm S}$ represents the separation between the neighboring clumps. 
We see significant variations in the ACF shape around its second peaks,
indicating different clump intervals at different $\theta$.
In other words, there is no coherence in the clump distributions. 
This feature can also be captured from close inspection of the lower-left panel of Figure \ref{2Dplots};
that is, we see a number of clumps in the radial directions but they are not equidistant.

To compare the ACFs of the 2D and 3D results more quantitatively,
we plot in Figure \ref{ACF-R} the ACFs calculated based on the 3D simulations (upper panel) 
and those based on the 2D simulations (lower panel), respectively.
In the upper panel the black line represents the azimuthally averaged ACF.
As expected, from comparing the density contours shown in Figure \ref{2Dplots}
we see that the clump width is thinner in the 3D simulations than in the 2D simulation. 
In addition, we find that the ACF of the 3D simulations
(red line in the upper panel) 
is similar to that of the 2D model (lower panel).
However, the azimuthally averaged ACF is 
much smoother, indicating that
the separation between the neighboring clumps is quite complicated,
as we have mentioned above (see also Figure \ref{ACF}).

\begin{figure}
  \begin{center}
    \includegraphics[width=0.8\linewidth]{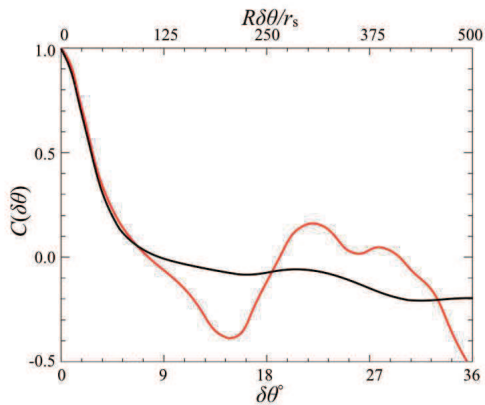}
  \end{center}
 \caption{As the upper panel of figure \ref{ACF-R}, but for the ACF as a function of $\delta\theta$. We fix the value of $R=740$ $r_{\rm S}$.}
  \label{ACF-theta}
\end{figure}

Finally, we plot the ACF as a function of $\delta\theta$ in Figure \ref{ACF-theta} based on
the matter density contour data displayed in the lower-left panel of Figure \ref{2Dplots}.
The half-width of the clumps in the azimuthal direction is estimated at 
$\ell_{\rm cl}^\theta/2 \sim 30$ $r_{\rm S}$,
which is a factor of $\sim$2 larger than that in the radial direction.
This gives a qualitative difference in the widths of the clumps, depending on the direction.

We admit that the ACF analysis is of limited use, since we can easily understand
from a quick look at the lower-left panel of Figure \ref{2Dplots} that
each clump is stretched out nearly (but not precisely) in the azimuthal direction and that
its length is much longer than $\sim 30$ $r_{\rm S}$, on the order of $R \sim$ several
hundreds of $r_{\rm S}$. The reason why much a shorter correlation length is found in the ACF analysis at fixed $R$ resides in the fact that each clump does not have a straight shape but shows a wavy structure.
This fact is essential when we discuss the obscuration of the central light by floating clumps (see the next section).

\section{Discussion}
\subsection{Formation Mechanism of Clumpy Outflows}
There are several physical mechanisms proposed for clump (or cloud) formation in gas flow; 
thermal instability (Field 1965), radiation-induced instability (Shaviv 2001), photo-bubble instability (Arons 1992), etc. When considering our particular case, we should keep in mind several key features confirmed by the 2D RHD simulations (see T13): \par
\noindent
(1) clumpy structure appears in the layer where upward radiation force overcomes downward gravity force,
\par\noindent
(2) a clump size is about one optical depth, 
\par\noindent
(3) there is an anti-correlation between the gas density and the absolute value of radiation force.
\par\noindent
(4) temperature variations of some clumps are neither monotonic increase nor monotonic decrease.

Not all but some similar features are confirmed in the present 3D simulation. 
The first one is obvious; we are now considering radiation-pressure-driven outflow.
The second feature can also be confirmed. The typical (radial) clump size and density are
$\ell_{\rm cl}^r \sim 30$ $r_{\rm S}\sim 10^{8}$cm and $\rho_{\rm cl} \sim 10^{-7}$ g cm$^{-3}$,
respectively (see the middle-left and lower-left panels of
Figure $\ref{2Dplots}$).
From these values we estimate the scattering optical depth 
($\tau_{\rm cl} = \kappa_{\rm es}\rho_{\rm cl}\ell_{\rm cl}^r$) to be a few.
Note that optical depth is expected to be 
independent of the black hole mass,
as long as the following scaling relations hold:
$\rho_{\rm cl}\propto M_{\rm BH}^{-1}$ and $\ell_{\rm cl}\propto M_{\rm BH}$ (see T13).
We do not have a clear correlation, however, between the matter density and radiation force,
as was demonstrated by T13 (see their Figure 4). 
Although the precise origin of clump formation in the 3D simulations requires further investigation,
we can at least say that
the combination of the Rayleigh--Taylor instability in the radiation-pressure-supported atmosphere
and a sort of radiation hydrodynamic instability seems to be involved with the clump formation.

The reason for asymmetric 3D shape of the clumps may be understood in terms of the anisotropic 
radiation field. That is, the radiation flux is highly super-Eddington in the radial direction,
while it is not in the angular direction. Hence, the Rayleigh-Taylor instability grows to
form a mushroom structure in the radial direction in the initial phase,
followed by the formation phase of individual clumps (with a typical size of one optical depth)
by a sort of radiation hydrodynamic instability (T14). 
Due to the differential rotation, however, such radially elongated clumps is soon stretched out
in the angular direction, thus forming a torn sheet structure. 

\subsection{Notes on the Numerical Resolution}
It is important to note that 
our 3D simulations produce larger clump sizes ($\sim 30 r_{\rm S}$) than those obtained
by the 
high-resolution 2D simulations ($\sim 10 r_{\rm S}$, see T13).
This is due partly to the 3D effects but
poor resolutions in 3D simulations may also affect the clump size.
In order to clarify this point, we performed 3D simulations with a grid spacing 
in the radial direction of 2.0 $r_{\rm S}$
(half of that of the simulations presented in the manuscript), 
while keeping the same spacing in the azimuthal direction, for a shorter simulation time (10 s), 
finding that the averaged clump size is slightly less ($\sim 25 r_{\rm S}$). 
We thus admit that the numerical resolution does affect the calculated clump size. Certainly, higher-resolution simulations are necessary to obtain a real clump size as future work. 

\subsection{Comparison with Observation}
In this section we discuss the observational implications in terms of three quantities:
variability timescale, ionization parameter, and volume filling factor.
\subsubsection{Variability timescale}
The clumpy nature of the outflow has been indicated thorough, e.g., X-ray spectral variations of 
luminous accretion flows, such as ULXs (e.g. Middleton et al. 2011; Pinto et al. 2016, 2017).
When clumps pass across the observer's line of sight towards the central bright region, 
the observed luminosity is expected to be temporarily reduced, since the scattering optical 
depth of each clump is a few or so (see section 4.1).
(The amount of reduction depends on the wavelengths of radiation, but we do not 
go into details in this discussion. The spectral variations are left as future work.)
Along this line, let us estimate the variation timescales. 

 From the simulation results we find that clumps are seen at distances greater than
 \begin{equation}
   R_{\rm cl} \sim 10^3 \: r_{\rm S}
   \sim 10^{9.5} \left(\frac{M_{\rm BH}}{10 M_\odot}\right) {\rm cm},
 \end{equation}
and the clump length in the azimuthal direction is 
\begin{equation}
  \ell_{\rm cl}^\theta \sim 10^2 \: r_{\rm S}
  \sim 10^{8.5} \left(\frac{M_{\rm BH}}{10 M_\odot}\right) {\rm cm}.
\end{equation}
from the ACF analysis. 
It might be thought that
the real clump lengths could be much longer, $\ell_{\rm cl}^\theta \sim 10^3$ $r_{\rm S}$
at longest, from close inspection of the density contours (see subsection 3.2).
But as a conservative estimate, we continue to use the value of $10^2$ $r_{\rm S}$.

The clump velocities are, on the other hand, of the order of 
$\sim 0.3 V_{\rm K}$,
where the Keplerian velocity is (see equation \ref{Kepler})
\begin{equation}
  V_{\rm K} = \sqrt{\frac{GM_{\rm BH}R_{\rm cl}^2}{r_{\rm cl}^3}} 
   \simeq 10^{8.6}
   \left(\frac{R_{\rm cl}}{10^3 r_{\rm S}}\right)^{-1/2}
    {\rm cm~s}^{-1},
\end{equation}
where we set $r_{\rm cl} \simeq \sqrt{2} R_{\rm cl}$ (i.e., $\theta = 45\deg$). 
Then, the variation timescale is estimated to be
\begin{equation}
  t_{\rm var} \sim \frac{\ell_{\rm cl}^\theta }{0.3 V_{\rm K}}
   \sim 2.5\left(\frac{M_{\rm BH}}{10 M_\odot}\right)
   \left(\frac{R_{\rm cl}}{10^3 r_{\rm S}}\right)^{1/2}
   \left(\frac{\ell_{\rm cl}^\theta }{10^2 \: r_{\rm S}}\right){\rm s},
\end{equation} 
In conclusion, the variability timescale could be several seconds for a $10 M_\odot$ black hole and 
could be even longer for a more massive black hole, since it is proportional to the black hole mass.

\begin{figure}
  \begin{center}
    \includegraphics[width=\linewidth]{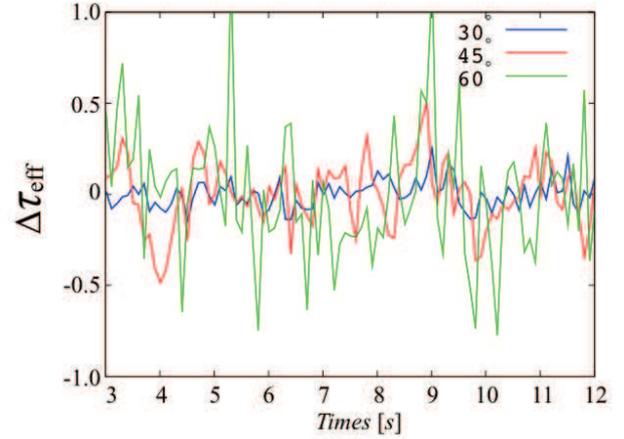}
  \end{center}
  \caption{Time variations of the effective optical depth with respect to the averaged one
for a variety of the angles measured by a distant observer at 
$\phi = 30^\circ, 45^\circ,$ and $60^\circ$ for a fixed $\theta = 45^\circ$.}
  \label{variation}
\end{figure}

We also attempt to estimate the variation timescales directly from the simulation data by 
calculating the time variations of effective optical depth 
($\tau_{\rm eff}$) by integrating the effective opacity 
($\sqrt{3 \kappa_{\rm es}\kappa_{\rm ff}}$ 
for $\kappa_{\rm es}\gg \kappa_{\rm ff}$) 
from the outer boundary of the simulation box to
$r=300 r_{\rm S}$. We have chosen the radial distance 
of 300 $r_{\rm S}$ since this layer crudely corresponds to the location
of the mean 
photosphere in the sense that the effective optical depth is around unity. 
In Figure \ref{variation}, we show the relative variations in the
effective optical depth from the average value, postulating that optical depth variations will lead to time changes in the flux because of absorption by floating clouds above the mean photosphere.
This is for the case that the observer is at infinite distance from the center in the directions of
$\phi = 30^\circ, 45^\circ,$ and $60^\circ$ for a fixed $\theta = 45^\circ$,
where $\phi (\equiv \tan^{-1} R/z)$ is the inclination angle.

Except during the first 3 s, the initial transient phase,
the optical depth varies on a timescale of several seconds. 
This would imply that the aperiodic fluctuation of the luminosity will be observed on this timescale.
The amplitudes of variations would roughly be 
$\sim 50$\%$-80$\% because of $|\Delta\tau_{\rm eff}| \sim 0.2-0.3$, 
depending on the viewing angle.
We admit, however, that this is a very crude estimation, and hence,
we eventually need a 3D radiation transfer calculation
to make the point clear.

This timescale is consistent with the variation timescale of a ULX estimated by Middleton et al. (2011).
That is, the mass of the central objects of the ULX is, at most, on the order of several tens of $M_\odot$.  
We can safely conclude that the central objects should be stellar-mass black holes.

Our result gives a significantly longer timescale than that estimated by T13 based on their 2D simulations. 
There are two main reasons for this discrepancy:
 (1) the typical length scale of each clump is systematically longer
 than the clump width from our 3D simulations, and (2) the rotation velocity turns out to be by some factor smaller than the 
local Keplerian value,
although T13 estimated the timescale by assuming Keplerian rotation.


\subsubsection{Photoionization parameter and volume filling factor}
In order to discuss the observational properties of the clumpy outflow, we further calculate two key parameters characterizing the outflow clumps: the photoionization parameter and the volume filling factor (see T13).
   
Let us first estimate the photoionization parameter, which is defined as
\begin{equation}
   \xi = L_{\rm x}/{n_{\rm cl} r^2}
\end{equation}
where $L_{\rm x}$ is the luminosity in the X-ray band and $n_{\rm cl}$ is the (average) gas number density of the clumps. The latter is related to the (average) optical depth for Thomson scattering of the clumps through
\begin{equation}
 \tau_{\rm cl} = n_{\rm cl} \sigma_{\rm T} \ell_{\rm cl},
 \label{taucl}
\end{equation}
where $\sigma_{\rm T}$ is the Thomson cross-section, and $\ell_{\rm cl}$ is the average size of the clumps; i.e., $\ell_{\rm cl} \sim \sqrt{\ell_{\rm cl}^r \ell_{\rm cl}^\theta}$. Then, the photoionization parameter is estimated to be
\begin{equation}
   \xi \sim 10^3 \tau_{\rm cl}^{-1}
       \left(\frac{L_{\rm x}}{0.1 L_{\rm E}}\right)
       \left(\frac{\ell_{\rm cl}}{30~r_{\rm S}}\right)
       \left(\frac{r}{10^3~r_{\rm S}}\right)^{-2}  {\rm erg~cm~s^{-1}}.
 \label{xi}
\end{equation}
Equation (\ref{xi}) indicates that clumpy outflows are mildly ionized, 
as long as 
$\ell_{\rm cl}\sim 30r_{\rm S}$ 
\citep{1982ApJS...50..263K}.

It is important to note that the ionization parameter 
does not explicitly depend on the black hole mass, 
although the X-ray luminosity, $L_{\rm x}$,
weakly depends on it; $L_{\rm x}\sim L_{\rm bol}$ 
for Galactic sources (with $M_{\rm BH} \sim 10 M_\odot$) 
while $L_{\rm x}\sim 0.1 L_{\rm bol}$ for AGNs 
(with $M_{\rm BH} \sim 10^8 M_\odot$). 
Here, $L_{\rm bol}$ is the bolometric luminosity. 
We assume $L_{\rm bol}\sim L_{\rm E}$ in the estimation 
given in equation (\ref{xi}).
We thus expect line absorption features to be observed, 
in agreement with observations.

Let us next estimate the volume filling factor defined as
\begin{equation}
  {\cal F}=\frac{V_{\rm cl}}{4\pi r_{\rm out}^3 /3},
  \label{vol_fill}
\end{equation}
where $V_{\rm cl} (= M_{\rm cl}/n_{\rm cl} m_{\rm p})$ 
is the volume occupied by the clumps, with $M_{\rm cl}$
being the total mass of the clumps,
and $r_{\rm out}$ is the size of the outflow region.
Most of the mass of the outflow is contained in the clump,
so that we find $M_{\rm cl} \sim \dot{M}_{\rm out} (r_{\rm out}/v_{\rm wind})$,
with $\dot{M}_{\rm out}$ being the outflow rate 
and $v_{\rm wind}$ being the radial wind velocity.
Since $n_{\rm cl}$ is given by equation (\ref{taucl}),
we can obtain the filling factor as
\begin{eqnarray}
 {\cal F}
  &\sim& 4.5\times 10^{-3}
  \tau_{\rm cl}^{-1}
  \left(\frac{\dot{M}_{\rm out}}{10L_{\rm E}/c^2} \right)
  \left(\frac{v_{\rm wind}}{0.1c} \right)^{-1}
  \nonumber\\
  && \times \left(\frac{\ell_{\rm cl}}{30r_{\rm S}} \right)
  \left(\frac{r_{\rm out}}{10^3r_{\rm S}} \right)^{-2}.
  \label{vol_fill2}
\end{eqnarray}
This is consistent with the filling factor estimated by T13,
because the outflow rate, the wind velocity, and the clump size are similar
to those in T13.
The volume filling factor is independent of the black hole mass
because $\dot{M}_{\rm out} \propto M_{\rm BH}$,
$\ell_{\rm cl} (\propto n_{\rm cl}^{-1}) \propto M_{\rm BH}$,
and $r_{\rm out} \propto M_{\rm BH}$.
Thus, our results are applicable to AGNs, 
although we employ a black hole of 10 $M_\odot$ in the simulation.
If we employ $r_{\rm out} \sim 10^5 r_{\rm S}$, 
the filling factor becomes around $4.5\times 10^{-7}$,
which is roughly consistent with observations of the BLR clouds 
of the AGNs at around $r \sim$ 1 pc 
($\sim10^5 r_{\rm S}$ for $M_{\rm BH} \sim 10^8 M_\odot$;
\cite{1997iagn.book.....P}).
More quantitative analyses are left as future work.
\\

This work is partially supported by MEXT/JSPS KAKENHI Grant Numbers
15K05036, 16K05309, and 17H01102 (KO), 17K05383 (SM), and 17K14260 (HRT).
Numerical computations were mainly carried out on a Cray XC30 at the Center for Computational Astrophysics, National Astronomical Observatory of Japan.      
This research was also supported by MEXT as a Priority Issue on Post-K computer (Elucidation of the Fundamental Laws and Evolution of the Universe) and JICFuS.

\end{document}